# Noise and Disturbance Compensation Approach for Linear Time-Invariant Plants*

Igor B. Furtat*,**

* Institute for Problems of Mechanical Engineering Russian Academy of Sciences, 61 Bolshoy ave V.O., St.-Petersburg, 199178, Russia (Tel: +7-812-321-47-66; e-mail: cainenash@mail.ru)
** ITMO University, 49 Kronverkskiy ave, Saint Petersburg, 197101, Russia.

**Abstract:** The algorithm with compensation of parametric uncertainties, external disturbances and measurement noises for linear time-invariant plants is designed. It is assumed, that the dimension of the noise can be equaled to the state vector dimension and the disturbance can be presented in any equation of the plant model. Analytical condition for algorithm feasibility is proposed. Simulation results show the efficiency of the proposed algorithm.

1. INTRODUCTION

There is a special attention of disturbance compensation in a theory and practice of automatic control. This is due to the fact that the technical and technological process are negatively influenced by external disturbances. The disturbance compensation approach is based on the design the control law such that the value of the control law is opposite to the value of the disturbance. This approach allows us to control of the plant without a significant increase of the control law magnitude. Currently, there are many methods for control with disturbance compensation. For example, the method of invariant ellipsoids (Polyak and Topunov, 2008), methods based on the internal disturbance model (Nikiforov, 1997), the methods of identification of sinusoidal signal parameters (Fedele and Ferrise, 2013), the method of imbedded systems (Bukov and Bronnikov, 2006), the universal controller method (Proskurnikov and Yakubovich, 2012), the auxiliary loop method (Tsykunov, 2010), etc.

The problem of control under disturbances becomes more complicated if the measured plant output contains noises. Noise measurements are the influence of the environment, measuring device characteristics (absolute and relative errors, accuracy class, the device type, etc.). For example (Guo et al., 2000, Chen, 2015), in multi-machine power systems external disturbance are caused by changing of resistance of transmission lines (short circuit, diurnal variation of the load, ice, etc.). Noises in multi-machine power systems are caused by the presence of measurement device errors in the measurements of the rotor speed, phase and EMF in each electrical generator, etc. (especially in faults).

Currently, the H∞-optimization (Sanchez-Peña and Sznaier, 1998), the auxiliary loop method (Tsykunov, 2010), methods of zooming in quantizer (Delchamps, 1989, Baillieul, 2002, Furtat et al., 2015) are used for control under measurement noises.

Note that in (Tsykunov, 2010) the author consider the plant model more general then model in (Sanchez-Peña and Sznaier, 1998). Additionally, differently from (Nikiforov, 1997) in (Tsykunov, 2010) external disturbances are assumed bounded only. However, in (Tsykunov, 2010) it is assumed that the dimension of the noise must be less than the dimension of the plant state vector; parametric and external disturbances may be present only in appropriate equations of the plant model. In addition, algorithm of (Tsykunov, 2010) contains high-gain observer. Therefore, the algorithm of (Tsykunov, 2010) has additional errors caused by high frequency signals in the disturbance.

In this paper, we design a novel algorithm for noise and disturbance compensation. In contrast to (Tsykunov, 2010), the design of proposed algorithm is based on results from (Furtat, 2016). The designed algorithm compensates the noise which dimension is equal to the dimension of state vector and compensates the disturbance which can be located in any equation of the plant model. Additionally, in contrast to (Tsykunov, 2010), the proposed algorithm allows one to obtain the analytical conditions for calculation of the controller parameters.

The rest of the paper is outlined as follows. In Section 2, the problem formulation is presented. In Sections 3, the noise compensation algorithm is designed. In Section 4, the noise

* The noise compensation algorithm (Section 3) was developed under support of RSF (grant 14-29-00142) in IPME RAS. The disturbance and noise compensation algorithm design (Section 4) was supported solely by the Russian Federation President Grant (No. 14.W01.16.6325-MD (MD-6325.2016.8)). The other research were partially supported by grants of Russian Foundation for Basic Research No. 14-08-01015, 16-08-00282, № 16-08-00686, Ministry of Education and Science of Russian Federation (Project 14.Z50.31.0031) and Government of Russian Federation, Grant 074-U01.

compensation algorithm is generalized on compensation of disturbance. Analytical condition for controller feasibility is proposed. In Section 5, simulation results illustrate effectiveness of the proposed scheme and confirm analytical results. Section 6 collects some conclusions. Finally, the proof of theorem is considered in Appendix A.

2. PROBLEM FORMULATION

Let a plant model be described by the following equation

$$\dot{x}(t) = Ax(t) + Bu(t) + Bf(x, u, t), \quad (1)$$

$$z(t) = x(t) + \xi(t), \quad (2)$$

where $x(t) \in R^n$ is a state vector, $u(t) \in R$ is a control, $f(x, u, t) \in R$ is an unmeasured disturbance signal depending on parametric uncertainties and external disturbances, $z(t) \in R^n$ is a measured signal, $\xi(t) = [\xi_1(t), ..., \xi_n(t)]^T$ is an unmeasured bounded noise and $\dot{\xi}(t)$ is bounded signal. The matrix $A \in R^{n \times n}$ is Hurwitz, $B = [b_1, ..., b_n]^T$ and the pair $(A, B)$ is controllable.

We assume that the function $f(t) = f(x, u, t)$ is given by

$$f(t) = c_{01}^T x(t) + c_{02} u(t) + c_{03} \varphi(t), \quad (3)$$

where $c_{01} \in R^n$, $c_{02} > -1$ and $c_{03} \in R$ are unknown vector and coefficients which belong to known compact set $\Xi$, signals $\varphi(t)$ and $\dot{\varphi}(t)$ are unknown and bounded.

It is necessary to design the algorithm ensuring the following goal

$$|x(t)| < \delta \text{ for } t > T, \quad (4)$$

where $\delta > 0$ is an accuracy, $T > 0$ is a transient time. Estimates on $\delta$ will be given in Theorem below. Here $|\cdot|$ is the Euclidean norm of a corresponding vector.

3. NOISE COMPENSATION ALGORITHM

Rewrite equation (2) in the form

$$z(t) = x(t) + \sum_{j=1}^{n} E_j \xi_j(t). \quad (5)$$

Here $E_j = [0, ..., 0, 1, 0, ..., 0]^T$ is a vector of corresponding dimension, where the $j$th component is equal to 1 and other components are equal to 0. Denote

$$\tilde{\xi}(t) = [\xi_1(t), ..., \xi_{i-1}(t), \xi_{i+1}(t), ..., \xi_n(t)]^T,$$
$$\tilde{E} = [E_1, ..., E_{i-1}, E_{i+1}, ..., E_n]^T,$$

and rewrite (5) as follows

$$z(t) = x(t) + \tilde{E}\tilde{\xi}(t) + E_i \xi_i(t). \quad (6)$$

Eliminate the $i$th equation in (6). To this end, multiply (6) by $\tilde{I}$, where $\tilde{I}$ is an $(n-1) \times n$ matrix obtained from the identity matrix of order $n$ by deleting the $i$th row. As a result, we have

$$\tilde{z}(t) = \tilde{I}x(t) + \tilde{\xi}(t), \quad (7)$$

where $\tilde{z}(t) = \tilde{I}z(t)$.

Differentiating (7) along the trajectories of system (1), we obtain

$$\dot{\tilde{z}}(t) = \tilde{I}Ax(t) + \tilde{I}Bu(t) + \tilde{I}Bf(t) + \dot{\tilde{\xi}}(t). \quad (8)$$

Expressing the variable $x(t)$ in (6) and substituting $x(t)$ into (8), we get

$$\dot{\tilde{z}}(t) = \tilde{I}Az(t) - \tilde{I}A\tilde{E}\tilde{\xi}(t) - \tilde{I}AE_i \xi_i(t) + \tilde{I}Bu(t) + \tilde{I}Bf(t) + \dot{\tilde{\xi}}(t). \quad (9)$$

Denote

$$\tilde{A} = \tilde{I}A\tilde{E}, \quad \tilde{A}_1 = \tilde{I}A, \quad \tilde{A}_2 = \tilde{I}AE_i, \quad \tilde{B} = \tilde{I}B,$$

and rewrite (9) in the form

$$\dot{\tilde{\xi}}(t) = \tilde{A}\tilde{\xi}(t) + \dot{\tilde{z}}(t) - \tilde{A}_1 z(t) - \tilde{B}u(t) - \tilde{B}f(t) + \tilde{A}_2 \xi_i(t). \quad (10)$$

Let the vector $\hat{\tilde{\xi}}(t)$ is an estimate of the vector $\tilde{\xi}(t)$. Define $\hat{\tilde{\xi}}(t)$ as follows

$$\dot{\hat{\tilde{\xi}}}(t) = \tilde{A}\hat{\tilde{\xi}}(t) + \dot{\tilde{z}}(t) - \tilde{A}_1 z(t). \quad (11)$$

Note, that the one of the implementations of (11) can be represented in the form

$$\hat{\tilde{\xi}}(t) = \int_0^t \left[\tilde{A}\hat{\tilde{\xi}}(s) - \tilde{A}_1 z(s)\right] ds + \tilde{z}(t) + \gamma, \quad (12)$$

where $\gamma = \hat{\tilde{\xi}}(0) - \hat{z}$, $\hat{z}$ is an estimate of $\tilde{z}(0)$.

Consider the error

$$e(t) = \tilde{\xi}(t) - \hat{\tilde{\xi}}(t), \quad (13)$$

where $e(t)$ defines the estimation quality of the vector $\tilde{\xi}(t)$ by using algorithm (12). Taking into account (10) and (11), differentiate $e(t)$:

$$\dot{e}(t) = \tilde{A}e(t) - \tilde{B}(u(t) + f(t)) + \tilde{A}_2 \xi_i(t). \quad (14)$$

It follows from equation (14) that the value of $e(t)$ depends on the values of $\xi_i(t)$ and $f(t)$. Moreover, $e(t)$ can be reduced if the control law $u(t)$ is chosen such that $u(t) = -f(t)$. Therefore, the next section is dedicated to design of the control law $u(t)$.

## 4. DISTURBANCE COMPENSATION ALGORITHM

Consider the *i*th equation of system (1). Multiplying (1) by the vector $E_i^T$, we get

$$\dot{x}_i(t) = E_i^T A x(t) + E_i^T B u(t) + E_i^T B f(t). \quad (15)$$

Express in (15) the disturbance signal *f(t)* in the form

$$f(t) = (E_i^T B)^{-1} [\dot{x}_i(t) - E_i^T A x(t) - E_i^T B u(t)] \quad (16)$$

It follows from (16) that the disturbance *f(t)* depends on unmeasured signals *x(t)* and $\dot{x}_i(t)$. Therefore, we will obtain the estimates of *x(t)* and $\dot{x}_i(t)$.

Let $\hat{x}(t)$ be an estimate of the vector *x(t)*. Taking into account (6) and (13), introduce the vector $\hat{x}(t)$ as follows

$$\hat{x}(t) = z(t) - \tilde{E}\hat{\xi}(t) = x(t) + \tilde{E}e(t) + E_i\xi_i(t). \quad (17)$$

Introduce the disturbance estimate $\hat{f}(t)$ in the form

$$\hat{f}(t) = (E_i^T B)^{-1} [\dot{\hat{x}}_i(t) - E_i^T A \hat{x}(t) - \alpha(p) E_i^T B u(t)] \quad (18)$$

where $\alpha(p)$ is a linear differential operator. The operator $\alpha(p)$ is needed for implementation of control law *u(t)*. Otherwise, the control system is not feasible if $\alpha(p) = 1$.

Introduce the disturbance compensation control law in the form

$$u(t) = -\hat{f}(t). \quad (19)$$

Substituting (18) into (19), we obtain

$$(1 - \alpha(p))u(t) = -(E_i^T B)^{-1} [\dot{\hat{x}}_i(t) - E_i^T A \hat{x}(t)] \quad (20)$$

For simplicity, let $\alpha(p) = 1 - \mu p$, where $\mu > 0$ is a design parameter. Rewrite control law (20) in the form

$$u(t) = -\frac{1}{\mu}(E_i^T B)^{-1}\left[\hat{x}_i(t) - E_i^T A \int_0^t \hat{x}(s)ds - \theta\right], \quad (21)$$

where $\theta = s(0) + \mu(E_i^T B)^{-1} u(0)$, *s(0)* is an estimate of $\hat{x}(0)$.

Introduce the following notations:

$$A_{21} = \frac{1}{\mu}(1+c_{02})A, \quad A_{22} = \frac{1}{\mu}(-I_n - I_n c_{02} + \mu B c_{01}^T + \mu A),$$

$$A_{23} = \frac{1}{\mu}B(1+c_{02})(E_i^T B)^{-1} E_i^T A \tilde{E}, \quad A_{42} = -\tilde{B}c_{01}^T,$$

$$A_{43} = \frac{1}{\mu}(1+c_{02})\left[\tilde{A} - \tilde{B}(E_i^T B)^{-1} E_i^T A \tilde{E}\right],$$

$$A_{44} = \frac{1}{\mu}(-(1+c_{02})I_n + \mu \tilde{A})$$

$$A_e = \begin{bmatrix} O_{n \times n} & I_n & O_{n \times (n-1)} & O_{n \times (n-1)} \\ A_{21} & A_{22} & A_{23} & O_{n \times (n-1)} \\ O_{(n-1) \times n} & O_{(n-1) \times n} & O_{(n-1) \times (n-1)} & I_{n-1} \\ O_{(n-1) \times n} & A_{42} & A_{43} & A_{44} \end{bmatrix}. \quad (22)$$

Here $O_{n \times l}$ is the zero matrix of dimension $n \times l$, $I_n$ is the identity matrix of order *n*.

**Theorem.** Let there exists the coefficient $\mu$ such that the matrix $A_e$ is Hurwitz. Then control system (12), (17) and (21) ensures goal (4).

Theorem will be proved in Appendix A.

**Corollary.** It follows from the proof of Theorem that the estimate of the value of $\delta$ can be defined as

$$\delta = \left(\lambda_{\min}^{-1}(P)\left[(V(0) - \rho^{-1}\chi^{-1}\bar{d}^2\right)\exp(\rho T) \right. \\ \left. + \rho^{-1}\chi^{-1}\bar{d}^2\right])^{0.5}. \quad (23)$$

Here

$$V = x_e^T(t) P x_e(t), \quad (24)$$

$x_e(t) = [x(t), \dot{x}(t), e(t), \dot{e}(t)]^T$, $P = P^T > 0$ is a solution of the equation $A_e^T P + P A_e = -Q$, $Q = Q^T > 0$, $R = Q - \chi P B_e B_e^T P$, $\chi > 0$, $R > 0$, $\rho = \lambda_{\min}(R)/\lambda_{\max}(P)$, $\bar{d} = \sup_{t \geq 0}|d(t)|$, $d(t) = [\dot{\varphi}(t), \xi_i(t), \dot{\xi}_i(t)]^T$,

$$B_e = \begin{bmatrix} O_{n \times 1} & O_{n \times 1} & O_{n \times 1} \\ B_{21} & B_{22} & B_{23} \\ O_{(n-1) \times 1} & O_{(n-1) \times 1} & O_{(n-1) \times 1} \\ B_{41} & B_{42} & B_{43} \end{bmatrix}, \quad (25)$$

$$B_{21} = Bc_{03}, \quad B_{22} = -\frac{1}{\mu}B(1+c_{02})(E_i^T B)^{-1} E_i^T A E_i,$$

$$B_{23} = -\frac{1}{\mu}B(1+c_{02})(E_i^T B)^{-1},$$

$$B_{41} = -\tilde{B}c_{03}, \quad B_{42} = -\frac{1}{\mu}\tilde{B}(1+c_{02})(E_i^T B)^{-1} E_i^T A E_i + \frac{1}{\mu}\tilde{A}_2,$$

$$B_{43} = \frac{1}{\mu}\tilde{B}(1+c_{02})(E_i^T B)^{-1}.$$

The ultimate bound of $\delta$ can be find as

$$\limsup_{t \to \infty} |x(t)| \leq \sqrt{\lambda_{\min}^{-1}(P)\rho^{-1}\chi^{-1}} \bar{d}. \quad (26)$$

**Remark 1.** It follows from (14) and the proof of Theorem that the value of $\delta$ significantly depends on $\xi_i(t)$. Therefore, for synthesis of the algorithm we can use the following recommendation. Assume that the vector *B* has nonzero components $b_l$ and $b_k$. Let the signals $\xi_k(t)$ and $\xi_l(t)$ have zero means and $\int_T^t |\xi_k(s)|ds \le \int_T^t |\xi_l(s)|ds$ for $t > T$. Thus, it is recommended to choose the $k$th equation in (15) for the algorithm synthesis.

**Remark 2.** Note, that only one integrator must be used for implementation of $\int_0^t [\tilde{A}\hat{\xi}(s) - \tilde{A}_1 z(s)]ds$ in (12). If the signals $\int_0^t \hat{\xi}(s)ds$ and $\int_0^t z(s)ds$ are implemented separately then the boundedness of $\int_0^t \xi(s)ds$ is required.

Let us illustrate given results on a numerical example.

## 5. EXAMPLE

Let model (1), (2) be presented in the form

$$\dot{x}(t) = \begin{bmatrix} -3 & 1 & 0 \\ -3 & 0 & 1 \\ -1 & 0 & 0 \end{bmatrix} x(t) + \begin{bmatrix} 1 \\ 1 \\ 3 \end{bmatrix} u(t) + \begin{bmatrix} 1 \\ 1 \\ 3 \end{bmatrix} f(x,u,t), \quad (27)$$

$$z(t) = x(t) + \xi(t).$$

The set of the possible values $\Xi$ is given by the following inequalities:

$$|c_{01}^1| \le 5, \ |c_{01}^2| \le 5, \ |c_{01}^3| \le 5, \ 0.5 \le c_{02} \le 2, \ |c_{03}| \le 1, \quad (28)$$

where $c_{01} = [c_{01}^1, c_{01}^2, c_{01}^3]^T$. In additional, we assume that $|\varphi(t)| \le 15$.

Design a control system. Let the third equation of (27) be the $i$th equation (see (15)). Then, $\tilde{I} = \begin{bmatrix} 1 & 0 & 0 \\ 0 & 1 & 0 \end{bmatrix}$. Therefore, $\tilde{A} = \begin{bmatrix} -3 & 1 \\ -1 & 0 \end{bmatrix}$, $\tilde{A}_1 = \begin{bmatrix} -3 & 1 & 0 \\ -3 & 0 & 1 \end{bmatrix}$. Let $\gamma = 0$. Introduce the noise estimation algorithm (12) as follows

$$\hat{\xi}(t) = \int_0^t \left( \begin{bmatrix} -3 & 1 \\ -1 & 0 \end{bmatrix} \hat{\xi}(s) - \begin{bmatrix} -3 & 1 & 0 \\ -3 & 0 & 1 \end{bmatrix} z(s) \right) ds + \tilde{z}(t),$$

where $\tilde{z}(t) = [z_1(t), z_2(t)]^T$.

Since $E_i = [0\ 0\ 1]^T$, then $E_i^T B = 3$ and $E_i^T A = [-1\ 0\ 0]$. Let $\theta = 0$. According to (21), the control signal is defined as

$$u(t) = -\frac{1}{3\mu}\left( \hat{x}_i(t) - [-1\ 0\ 0]\int_0^t \hat{x}(s)ds \right), \quad (29)$$

where, according to (17), $\hat{x}(t) = z(t) - \begin{bmatrix} 1 & 0 & 0 \\ 0 & 1 & 0 \end{bmatrix}^T \hat{\xi}(t)$.

Note that the matrix $A_e$ is Hurwitz for plant (27) with uncertainties (28) and control law (29) if $\mu \le 0.0503$. Fig. 1 shows the relation of the maximum value of the real part of the eigenvalues of the matrix $A_e$ (max(Re(eig($A_e$)))) on parameter $\mu$.

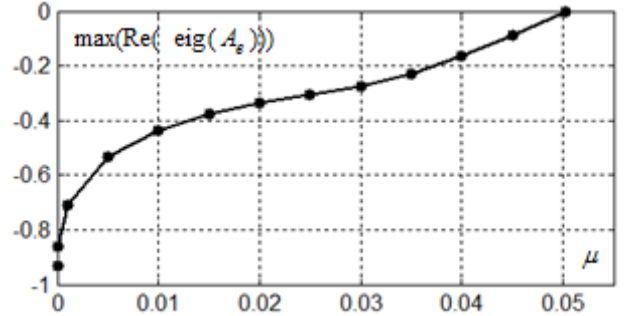

Fig. 1. The relation of the maximum value of the real part of the eigenvalues of the matrix $A_e$ (max(Re(eig($A_e$)))) on parameter $\mu$

Choose $\mu = 0.001$ in (29). Let $x(0) = [1\ 1\ 1]^T$ and

$$f(t) = 5x_1(t) + 5x_2(t) + 5x_3(t) + 0.5u(t) + 5$$
$$+ 5\sin 0.7t + \frac{1}{0.01p+1}d_1(t),$$

$$\xi_1(t) = 1 + 10\sin 3t + \frac{1}{0.001p+1}d_2(t), \quad (30)$$

$$\xi_2(t) = -2 + 7\cos 3t + \frac{1}{0.001p+1}d_3(t),$$

$$\xi_3(t) = 0.01\sin 0.8t + \frac{2}{p+1}d_4(t).$$

In (30) the signals $d_1(t)$, $d_2(t)$, $d_3(t)$ and $d_4(t)$ are piecewise constant with normally distributed random values with zero means, unit variances and sample time of 0.07 s, 0.01 s, 0.03 s and 0.1 s respectively. Additionally, $|d_1(t)| \le 5$, $|d_2(t)| \le 10$, $|d_3(t)| \le 12$ and $|d_4(t)| \le 0.05$. In Fig. 2 the transients of $x(t)$ are shown.

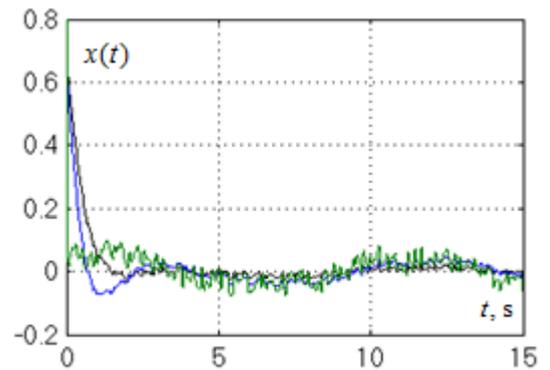

Fig. 2. The transients of $x(t)$ under disturbance and noise of (30)

Now consider the case of quantized measurements and disturbance in the form

$$f(t) = 5x_1(t) - 5x_2(t) + 5x_3(t) + 2u(t) + 7$$
$$+ 3\sin 2t + \frac{1}{0.01p + 1}d_1(t),$$
$$z_1 = \frac{1}{0.001p + 1}q_1(x_1), \quad z_2 = \frac{1}{0.001p + 1}q_2(x_2), \quad (31)$$
$$z_3 = \frac{1}{0.001p + 1}q_3(x_3),$$

where $q_1$, $q_2$ and $q_3$ are quantization functions with the quantization interval of 0.5, 1.3 and 0.05 respectively. In (Delchamps, 1989, Baillieul, 2002) it is noted that the quantization can be represented as an additive measurement noise. In Fig. 3 the transients of $x(t)$ are shown.

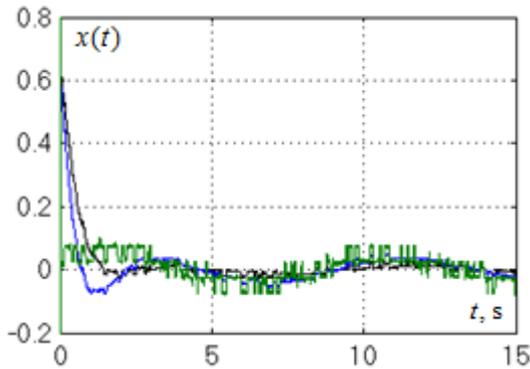

Fig. 3. The transients of $x(t)$ under disturbance and noise of (31)

In Fig. 2, 3 the signal $x_1(t)$ is a black curve, $x_2(t)$ is a blue curve and $x_3(t)$ is a green curve. It follows from Fig. 2, 3 that the control system compensates disturbances and measurement noises after 3 (s) with the accuracies of 0.1 and 0.05 respectively.

## 6. CONCLUSIONS

The disturbance and noise compensation algorithm for linear plants with unknown parameters is designed. In contrast to existing results, the proposed algorithm compensates the noise with dimension is equal to the dimension of the plant state vector. Moreover, the proposed algorithm does not use observer and disturbance can be presented in any equation of the plant model. Also, differently from (Tsykunov, 2010), we obtain conditions that allow us to calculate the design parameters in the algorithm. It is shown that the transient quality in steady state essentially depends on the noise component corresponding to equation for disturbance compensation. The simulation results confirmed the results of analytical calculations and the effectiveness of the control system.

## APPENDIX A

**Theorem proof.** Taking into account (16) and (18), rewrite (19) in the form

$$u(t) = -c_{01}^T x(t) - c_{02}u(t) - c_{03}\varphi(t)$$
$$- (E_i^T B)^{-1}\left[\dot{\hat{x}}_i(t) - \dot{x}_i(t)\right] + (E_i^T B)^{-1} E_i^T A[\hat{x}(t) - x(t)] \quad (A.1)$$
$$+ (\alpha(p) - 1)u(t).$$

Substituting $\alpha(p) = 1 - \mu p$ into (A.1), rewrite (A.1) as follows

$$(1 + c_{02} + \mu p)u(t)$$
$$= -c_{01}^T x(t) - c_{03}\varphi(t) \quad (A.2)$$
$$- (E_i^T B)^{-1}[\hat{x}_i(t) - \dot{x}_i(t)] + (E_i^T B)^{-1} E_i^T A[\hat{x}(t) - x(t)].$$

Taking into account (A.2) and (17), consider the following expression

$$(1 + c_{02} + \mu p)(u(t) + f(t)) = \mu p(c_{01}^T x(t) + c_{03}\varphi(t))$$
$$- (1 + c_{02})(E_i^T B)^{-1}(\dot{\xi}_i(t) - E_i^T A[\tilde{E}e(t) + E_i\xi_i(t)]). \quad (A.3)$$

Multiplying (1) by the operator $(1 + c_{02} + \mu p)$ and taking into account (A.3), rewrite (1) in the form

$$(1 + c_{02} + \mu p)(\dot{x}(t) - Ax(t))$$
$$= B\big[\mu p(c_{01}^T x(t) + c_{03}\varphi(t)) \quad (A.4)$$
$$- (1 + c_{02})(E_i^T B)^{-1}(\dot{\xi}_i(t) - E_i^T A[\tilde{E}e(t) + E_i\xi_i(t)])\big].$$

Rewrite (A.4) as follows

$$\ddot{x}(t) = \frac{1}{\mu}\big(-I_n - c_{02}I_n + \mu B c_{01}^T + \mu A\big)\dot{x}(t)$$
$$+ \frac{1}{\mu}(1 + c_{02})Ax(t) + B c_{03}\dot{\varphi}(t) \quad (A.5)$$
$$- \frac{1}{\mu}(1 + c_{02})B(E_i^T B)^{-1}$$
$$\times (\dot{\xi}_i(t) - E_i^T A\tilde{E}e(t) + E_i^T AE_i\xi_i(t)).$$

Denote $\eta_1(t) = x(t)$, $\eta_2(t) = \dot{x}(t)$. Taking into account (22) and (25), transform differential equation (A.5) to the following system

$$\dot{\eta}_1(t) = \eta_2(t),$$
$$\dot{\eta}_2(t) = A_{21}\eta_1(t) + A_{22}\eta_2(t) + A_{23}e(t) \quad (A.6)$$
$$+ B_{21}\dot{\varphi}(t) + B_{22}\xi_i(t) + B_{23}\dot{\xi}_i(t).$$

Multiplying (1) by the operator $(1 + c_{02} + \mu p)$ and taking into account (A.3), rewrite (14) in the form

$$(1 + c_{02} + \mu p)(\dot{e}(t) - \tilde{A}e(t)) =$$
$$- \tilde{B}\mu p(c_{01}^T x(t) + c_{03}\varphi(t))$$
$$+ \tilde{B}(1 + c_{02})(E_i^T B)^{-1} \quad (A.7)$$
$$\times (\dot{\xi}_i(t) - E_i^T A[\tilde{E}e(t) + E_i\xi_i(t)]) + \tilde{A}_2\xi_i(t).$$

Rewrite (A.7) as follows

$$\ddot{e}(t) = \frac{1}{\mu}\big(-I_{n-1} - c_{02}I_{n-1} + \mu\tilde{A}\big)\dot{e}(t)$$
$$+ \frac{1}{\mu}\Big[(1 + c_{02})\tilde{A} - \tilde{B}(1 + c_{02})(E_i^T B)^{-1} E_i^T A\tilde{E}\Big]e(t)$$
$$- \tilde{B}c_{01}^T \dot{x}(t) - \tilde{B}c_{03}\dot{\varphi}(t) \quad (A.8)$$
$$+ \frac{1}{\mu}\tilde{B}(1 + c_{02})(E_i^T B)^{-1}(\dot{\xi}_i(t) - E_i^T AE_i\xi_i(t))$$
$$+ \frac{1}{\mu}\tilde{A}_2\xi_i(t).$$

Denote $\sigma_1(t) = e(t)$, $\sigma_2(t) = \dot{e}(t)$. Taking into account (22) and (25), transform (A.8) to the following system

$$\dot{\sigma}_1(t) = \sigma_2(t),$$
$$\dot{\sigma}_2(t) = A_{43}\sigma_1(t) + A_{44}\sigma_2(t) + A_{42}\dot{x}(t) \quad (A.9)$$
$$+ B_{41}\dot{\varphi}(t) + B_{42}\xi_i(t) + B_{43}\dot{\xi}_i(t).$$

Taking into account (22), (25), (A.6) and (A.9), consider the equation

$$\dot{x}_e(t) = A_e x_e(t) + B_e d(t). \quad (A.10)$$

Let there exists the positive coefficient $\mu$ such that the matrix $A_e$ is Hurwitz. Since $d(t)$ is bounded, then $x_e(t) = [x(t), \dot{x}(t), e(t), \dot{e}(t)]^T$ is ultimate bounded. Thus, the signals $z(t)$ and $\dot{z}(t)$, and hence $\tilde{z}(t)$ and $\dot{\tilde{z}}(t)$, are bounded from (5) and boundedness of $\xi(t)$ and $\dot{\xi}(t)$. Therefore, the signals $\hat{x}(t)$ and $\dot{\hat{x}}(t)$ are bounded from (17). It follows from (A.2), boundedness of $\varphi(t)$ and $c_{02} > -1$ that the signal $u(t)$ is bounded. The signal $\int_0^t \hat{x}(s)ds$ is bounded from (21).

The signal $\hat{\xi}(t)$ is bounded from (13). Therefore, the function $\int_0^t [\tilde{A}\hat{\xi}(s) - \tilde{A}_1 z(s)]ds$ is bounded from (12). Consequently, all signals are bounded in the closed-loop system.

Let us find the ultimate bound of $x(t)$. To this end, we consider Lyapunov function (24). Calculate the derivative of (24) along the trajectories of system (A.10):

$$\dot{V} = -x_e^T(t)Qx_e(t) + 2x_e^T(t)PB_e d(t). \quad (A.11)$$

Consider the following upper bound

$$2x_e^T(t)B_e d(t) \leq \chi x_e^T(t)PB_e B_e^T Px_e(t) + \chi^{-1}\bar{d}^2.$$

Estimate (A.11) in the form

$$\dot{V} \leq -x_e^T(t)Rx_e(t) + \chi^{-1}\bar{d}^2. \quad (A.12)$$

Transform (A.12) as follows

$$\dot{V}(t) \leq -\rho V(t) + \chi^{-1}\overline{d}^{\,2}. \quad (A.13)$$

Solving inequality (A.13) w.r.t. $V(t)$, we obtain

$$V(t) \leq \left(V(0) - \rho^{-1}\chi^{-1}|d|^2\right)\exp(-\rho t) + \rho^{-1}\chi^{-1}\overline{d}^{\,2}. \quad (A.14)$$

Taking into account (24), rewrite (A.14) as follows

$$\lambda_{\min}(P)|x|^2 \leq \left(V(0) - \rho^{-1}\chi^{-1}\overline{d}^{\,2}\right)\exp(-\rho t) + \rho^{-1}\chi^{-1}\overline{d}^{\,2}. \quad (A.15)$$

Estimates (23) and (26) follows from inequality (A.15).

Let us show that there exist the coefficient $\mu$ such that the matrix $A_e$ will be Hurwitz. Consider the case when $\mu \to 0$. Therefore, $\alpha(p) \to 1$. Taking into account (16)-(18) and (A.1), rewrite (18) as follows

$$\hat{f}(t) = f(t) + \left(E_i^{\mathrm{T}}B\right)^{-1}\dot{\xi}_i(t) - \left(E_i^{\mathrm{T}}B\right)^{-1}E_i^T A\left[\tilde{E}e(t) + E_i\xi_i(t)\right] \quad (A.16)$$

for $\mu \to 0$. Taking into account (19) and (A.16), rewrite (14) in the form

$$\dot{e}(t) = \left(\tilde{A} - \tilde{B}\left(E_i^{\mathrm{T}}B\right)^{-1}E_i^{\mathrm{T}}A\tilde{E}\right)e(t) + \tilde{A}_2\xi_i(t) + \tilde{B}(E_i^{\mathrm{T}}B)^{-1}\dot{\xi}_i(t) - \tilde{B}(E_i^{\mathrm{T}}B)^{-1}(E_i^{\mathrm{T}}A)E_i\xi_i(t). \quad (A.17)$$

If $A_{43}$ is Hurwitz, then $\tilde{A} - \tilde{B}\left(E_i^{\mathrm{T}}B\right)^{-1}E_i^{\mathrm{T}}A\tilde{E}$ is Hurwitz. Therefore, $e(t)$ is ultimate bounded. Taking into account (19) and (A.16), rewrite (1) in the form

$$\dot{x}(t) = Ax(t) - B(E_i^{\mathrm{T}}B)^{-1}\dot{\xi}_i(t) + B(E_i^{\mathrm{T}}B)^{-1}(E_i^{\mathrm{T}}A)E_i\xi_i(t) + B\left(E_i^{\mathrm{T}}B\right)^{-1}E_i^{\mathrm{T}}A\tilde{E}e(t). \quad (A.18)$$

Since $A$ is Hurwitz, $\xi_i(t)$, $\dot{\xi}_i(t)$ and $e(t)$ are bounded, then $x(t)$ is ultimate bounded. Therefore, there exist small enough coefficient $\mu$ such that the signals $x(t)$ and $e(t)$ are ultimate bounded.